\documentclass{emulateapj}
\usepackage{apjfonts}

\newcommand{\EXO}{\mbox{EXO 0748-676}}

\newcommand{\Msun}{\ensuremath{M_\odot}}

\shorttitle{Rotational Broadening of NS Spectral Features} 
\shortauthors{Chang et al.}

\begin{document}

\title{Rotational Broadening of Atomic Spectral Features from Neutron Stars} 

\author{Philip Chang\altaffilmark{1,2}, Sharon Morsink\altaffilmark{3},
Lars Bildsten\altaffilmark{1,4}, and Ira Wasserman\altaffilmark{5}}
\altaffiltext{1} {Department of Physics, Broida Hall, University of
California, Santa Barbara, CA 93106}
\altaffiltext{2} {Miller Fellow: Department of Astronomy, University of
California, Berkeley, CA 94720; pchang@astro.berkeley.edu}
\altaffiltext{3}{Theoretical Physics Institute, Department of Physics,
University of Alberta, Edmonton, AB, T6G 2J1, Canada; email:
morsink@phys.ualberta.ca} 
\altaffiltext{4}{Kavli Institute for
Theoretical Physics, Kohn Hall, University of California, Santa
Barbara, CA 93106, USA; email: bildsten@kitp.ucsb.edu}
\altaffiltext{5}{Center for Radiophysics and Space Research, Cornell
University, Ithaca, NY 14853; ira@astro.cornell.edu}

\begin{abstract}

  The discovery of the first gravitationally redshifted spectral line
from a neutron star (NS) by Cottam, Paerels and Mendez has triggered
theoretical studies of the physics of atomic line formation in NS
atmospheres. Chang, Bildsten and Wasserman showed that the hydrogenic
Fe H$\alpha$ line formed above the photosphere of a bursting NS is
intrinsically broad. We now include rotational broadening within
general relativity and compare the resulting profile to that observed
during Type I bursts from EXO 0748-676. We show that the fine
structure splitting of the line precludes a meaningful constraint on
the radius. Our fitting of the data show that the line forming Fe
column is ${\rm log}_{10} (N_{\rm Fe, n=2}/{\rm
cm^{-2}})=17.9_{-0.42}^{+0.27}$ and gravitational redshift $1+z
=1.345_{-0.008}^{+0.005}$ with 95\% confidence. We calculate the
detectability of this spectral feature for a large range of spins and
inclinations assuming that the emission comes from the entire
surface. We find that at 300 (600) Hz only 10-20\% (5-10\%) of NSs
would have spectral features as deep as that seen in EXO 0748-676.

\end{abstract}


\section{Introduction}

The observation of gravitationally redshifted atomic absorption lines
from the surface of the neutron star (NS) in EXO 0748-676 (Cottam,
Paerels \& Mendez 2002; hereafter CPM) provides a new constraint on
the nuclear equation of state. By adding spectra of 28 Type I X-ray
bursts observed with XMM-Newton, CPM detected an absorption line at 13
\AA\ during the peak of the burst, which moved to 13.75 \AA\ as the
surface temperature declined.  They associated these features with
hydrogen-like and helium-like Fe $n=2\rightarrow 3$ transitions,
implying a gravitational redshift of $1+z\approx 1.35$. Motivated by
these observations, Bildsten, Chang \& Paerels (2003; hereafter BCP)
explained why Fe can be present above the photosphere. They also
discussed the microphysics of the resonant Fe H$\alpha$ line, showing
that it is Stark broadened because of the large photospheric
densities, $\rho \approx 0.1-1\,{\rm g\,cm}^{-3}$ (Paerels 1997) and
confirmed London, Taam \& Howard's (1986) conclusion that the
radiation field determines the Fe ionization balance.

Chang, Bildsten \& Wasserman (2005, hereafter CBW) recently amplified
these points and performed a self-consistent radiative transfer and
statistical equilibrium calculation. They assumed that the Fe
H$\alpha$ line forming region is a thin layer above the continuum
photosphere, as expected from BCP's accretion-spallation scenario. The
H$\alpha$ line is a scattering dominated resonance line with a fine
structure splitting of 20.7 eV comparable to the observed width and
the scale of rotational broadening for a 10 km NS rotating at 45 Hz of
24 eV.  The lack of pulsations from this NS during accretion leads us
to estimate that $B<1-2\times 10^9\,{\rm G}$ needed for Zeeman splitting
to play a role in the formation of the Fe H$\alpha$ line (Loeb 2003).
CBW calculated the equivalent width (EW) as a function of Fe column,
$N_{\rm Fe}$, for varying effective temperatures and pressures in the
thin line-forming region. To match the observed EW required, a total
iron column of $N_{\rm Fe} \approx 1-3 \times 10^{20}\,{\rm cm}^{-2}$
was needed, higher than predicted by BCP by a factor of a few.

  In this Letter, we rotationally broaden the energy and angle
dependent line profiles of CBW and compare them to the
observations. We explain our method of calculating the rotational
broadening in \S 2 and show that the Schwarzschild+Doppler (S+D)
approximation (Miller \& Lamb 1998; Braje, Romani, \& Rauch 2000) is
extremely accurate in reproducing the line profiles from the full
general relativistic calculation, even up to 300 Hz.  We fit the data
in \S 3 and show that the intrinsically broad line profile prohibits
any meaningful constraint on the NS radius if the 44.7 Hz burst
oscillation seen by Villarreal \& Strohmayer (2004) is the spin
frequency.  At slow spins or at low inclinations, the line profile
shape is set by both rotational broadening and fine structure
splitting, making it absolutely necessary to use realistic line
profiles when modeling the lines in \EXO\ or any other favorably
oriented (and hence detectable) rapidly rotating NS.  We confirm the
importance of the intrinsic line profile in determining NS parameters
by constrasting it with a simple line profile. We close \S 3 by
accurately determining the redshift and the amount of Fe in the upper
atmosphere.  We conclude in \S 4 by summarizing our findings and
discussing the likelihood that lines like that seen in \EXO\ will be
found in the more rapidly rotating NSs.

\section{Rotational Broadening}

We compute the impact of rotational broadening on CBW's intrinsic line
profiles using two independent methods: a fully relativistic method
that employs ray-tracing in the spacetime of a rotating NS and an
approximate method which retains the most important relativistic
effects to lowest order in $v/c$.

In the fully relativistic approach, the spacetime metric for rotating
NSs with various spin frequencies, masses and equations of state are
computed numerically as described in Cadeau, Leahy \& Morsink
(2005a). Once the (non-spherical) surface of the NS has been located,
geodesics connecting the surface to the observer are computed.  We
discretize the surface into small patches and calculate
the zenith angle $\alpha$ between the local normal and initial photon
direction in the locally comoving frame for each patch.  We also
calculate the solid angle $d\Omega$ subtended by the patch as seen by
the observer and the redshift $z_*$.  The redshift, $z_*$ includes
both the gravitational redshift and Doppler-like terms resulting from
rotation.  Since the intensity, $I$, is both a function of energy and
$\cos\alpha$, the observed flux due to the patch is
\begin{equation}
dF_{\nu} = (1+z_*)^{-3}I(\nu(1+z_*),\cos\alpha) {d\Omega}.
\end{equation}
The observed spectrum is found by summing over all visible
patches. The technical details of the fully relativistic computational
method are presented by Cadeau et~al. (2005b).

We also use the simpler S+D approximation (Miller \& Lamb 1998; Braje,
Romani, \& Rauch 2000; also see Cadeau et al 2005a and Poutanen \&
Gierlinski 2003) to independently compute the spectrum. In the S+D
approximation the metric is approximated by the non-rotating
Schwarzschild metric, and appropriate Doppler factors correct for the
rotation. To lowest order in $v/c$, the redshift factor reduces to
$(1+z_*) = (1+z)(1-\delta) $ where $\delta = \beta {\sin \zeta \sin
\alpha \sin i}$, $\beta = 2 \pi R \nu_{\rm s}/(c\sqrt{1-2GM/(Rc^2)})$
is the scale of rotational broadening, $\nu_{\rm s}$ is the observed
spin of the NS, and $z = (1-2GM/(Rc^2))^{-1/2} -1$. The angle $\zeta$
is the azimuthal angle of the emitting patch about the vector pointing
from the center of the star towards the observer. This angle is zero
if the patch is in the plane defined by the spin axis and this vector.

\begin{figure}
\plotone{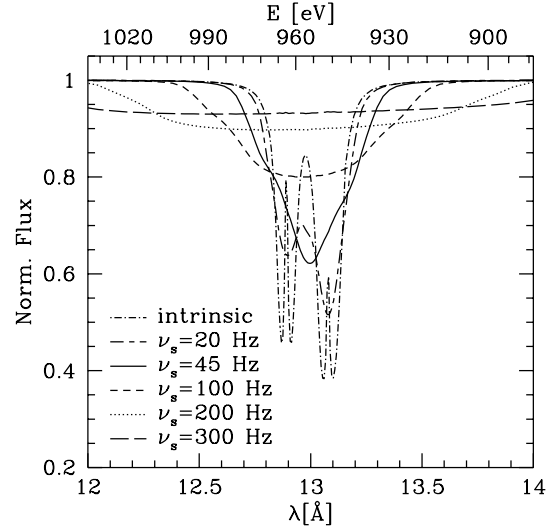}
\caption{Line profiles from an exact GR calculation (thick lines) and
  the S+D approximation (thin lines).  We plot four different models
  with $\nu_{\rm s}\sin i = 45$ Hz (solid lines), $\nu_{\rm s}\sin i =
  100$ Hz (short dashed lines), $\nu_{\rm s}\sin i = 200$ Hz (dotted
  lines), and $\nu_{\rm s} = 300$ Hz (long dashed lines).  The
  intrinsic line profile (dot-dashed line) is the same for these
  models.  We also plot the line profile for $\nu_{\rm s}\sin i = 20$
  Hz (long-short dashed line) to illustrate the line profile as it
  changes from being dominated by fine structure splitting to being
  dominated by rotational broadening.}\label{fig:Sharon}
\end{figure}

A further simplification is the use of the Beloborodov (2002)
approximation, $ 1-\cos\alpha = (1+z)^{-2} (1-\cos\psi), $ where
$\psi$ is the bending angle.  This allows the total observed flux to
be written as the integral over all possible initial photon zenith and
azimuthal angles,
\begin{equation}
F_\nu = (1+z)^{-1} \left(\frac {R}{d}\right)^2\int d\zeta \int
d(\cos\alpha) \cos\alpha \frac{
I\left(\nu(1+z_*),\cos\alpha\right)}{(1-\delta)^3},
\end{equation}
where $R$ is the NS radius and $d$ is the distance.  The results for
emission from the entire NS surface with CBW's intrinsic line profile
of the fully relativistic calculation and the S+D approximation are
shown in Figure \ref{fig:Sharon}.  They are in excellent agreement and
the tiny deviations are mainly due to our neglect of relativistic
aberation.  The linear change to the metric in $v/c$ due to frame
dragging (Hartle 1967, Hartle \& Thorne 1968) suggests observable
effect on the line at spin frequencies near 300 Hz as $\beta \sim
0.1$. However, the frame-dragging term enters into the line broadening
calculation as a second order effect in $v/c$, preserving the validity
of the simple S+D approximation. Only at very high spin frequencies
does frame dragging have the potential of producing observable effects
(Bhattacharyya et al 2004).

Of greater relevance is the relative size of the fine structure
splitting and the scale of rotational broadening.  At 45 Hz, the scale
of rotational broadening is 24 eV for a 10 km NS, which is similar to
the fine structure splitting of 20.7 eV for the Fe H$\alpha$ line.  At
slower and faster rotation rates, fine structure splitting dominates
rotational broadening and vice versa.  Hence the line profile can
change dramatically near 45 Hz.  We illustrate this effect in Figure
\ref{fig:Sharon} where we plot the line profile for $\nu_{\rm s}\sin i
= 20$ Hz (long-short dashed line), which is dominated by fine structure
splitting.

\section{Fits to EXO 0748-676 and Implications}

\begin{figure}
\epsscale{1.0} \plotone{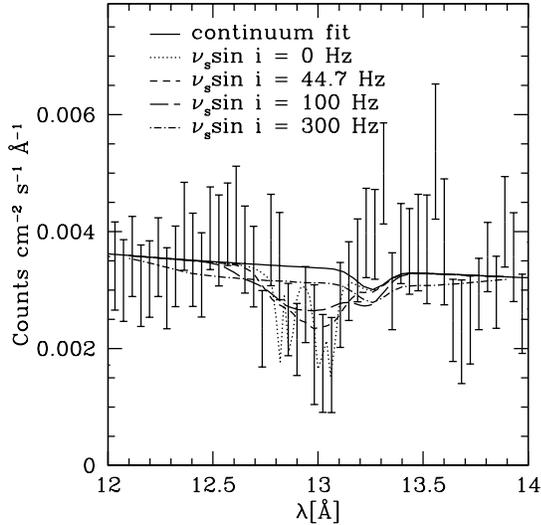}
  \caption{Line profiles at $R$ = 10 km for several different models
    compared to the data. We plot best fitting line profiles for
    $\nu_{\rm s} = 0$ Hz (dotted line), 44.7 Hz (short-dashed line),
    100 Hz (long-dashed line), and 300 Hz (dot-dashed line) for $R\sin
    i = 10$ km. We also show the continuum model of CPM (solid line)
    and the data (one sigma error bars).}
  \label{fig:fitSpectra}
\end{figure}

The simplicity and accuracy of the S+D approximation allows for a
rapid and detailed exploration of parameter space. We assume emission
from the entire surface.  Emission concentrated near the spin axis
would also give narrow line profiles (Bhattacharyya, Miller \& Lamb
2004), but we do not consider this possibility. We fit our models to
the co-added spectra and continuum model from CPM. We focus on a
narrow range between 12\AA\ and 14\AA\ and plot these points in
Figure \ref{fig:fitSpectra}. The 13.25 \AA\ feature in the continuum
fit is due to a Ne IX resonant transition in the circumstellar model.
Taking the continuum model as the background flux, we overlay some
models for $\nu_{\rm s} = 0$, $45$, 100 and 300 Hz for $R=10$ km.  To
set the scale of Stark broadening, we take a background density of
$n_p = 10^{23}\,{\rm cm}^{-3}$, which is appropriate for a uniform Fe
abundance above the continuum photosphere (CBW).  From these models,
large spins i.e. $\nu_{\rm s}\sin i = 300$ Hz are ruled out.
Secondly, the intrinsically broad line profile which is similar to the
width of the observed line, prohibits a meaningful limit on $R$ for
$\nu_{\rm s} = 44.7$ Hz.

We also performed a detailed exploration of parameter space. Using the
data between 12.6 \AA\ and 13.3 \AA, we overlay our line profiles on
the continuum model to test the goodness of fit.  Varying $\nu_{\rm
s}\sin i (R/10\,{\rm km})$, 1+z, and $N_{\rm Fe, n=2}$, the best fit
gives a $\chi^2 = 11.01$ for 17 degrees of freedom.  By itself, the
continuum model gives $\chi^2 = 24.9$. The best fit parameters are
$\nu_{\rm s}\sin i (R/10\,{\rm km}) = 32 \pm 19\,{\rm Hz}$, $1 + z =
1.345 \pm 0.004$, and ${\rm log}_{10} (N_{\rm Fe, n=2}/{\rm cm^{-2}}) =
17.9 \pm 0.2$. The errors indicated are one sigma values derived from
the covariance matrix.

\begin{figure}
\epsscale{1.0} \plotone{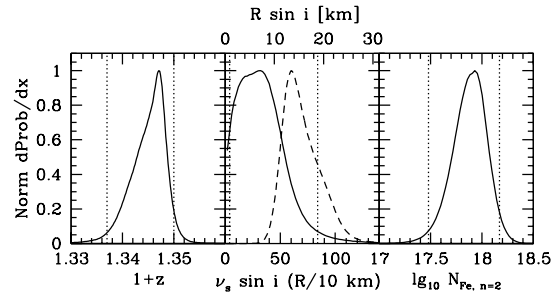} 
\caption{Marginalized probability distribution functions for $1+z$,
  $\nu_{\rm s}\sin i (R/10\,{\rm km})$ and $N_{\rm Fe, n=2}$
  normalized so that their peak is unity.  For the marginalized PDF of
  $\nu_{\rm s}\sin i (R/10\,{\rm km})$, we also display on the top
  axis the $R\sin i$ for $\nu_{\rm s} = 44.7$ Hz.  In addition, we
  plot the marginalized PDF of $\nu_{\rm s}\sin i (R/10\,{\rm km})$
  for a black line profile with fixed EW of $0.1$ \AA\ (dashed line).
  We also show the 95\% (dotted line) central confidence intervals for
  each parameter.}
\label{fig:likelihood}
\end{figure}

To develop a more detailed understanding of parameter space, we
marginalize our three parameter joint probability distribution
function (PDF) over two parameters at a time and plot the resulting
one dimensional PDF for each parameter in Figure \ref{fig:likelihood}.
The 68\%, 95\%, and 99\% central confidence intervals for the
intrinsic line profiles for $\nu\sin i (R/10\,{\rm km})$ ($R\sin i$
for $\nu_{\rm s} = 44.7\,{\rm Hz}$) are $11-52\,{\rm Hz}$
($0.25-11\,{\rm km}$), $4-84\,{\rm Hz}$ ($0.09 - 19\,{\rm km}$), and
$2-122\,{\rm Hz}$ ($<27\,{\rm km}$) respectively.  For the best
measured value of $1+z=1.345$, a 1.4 \Msun\ NS with R = 9.2 km falls
within these limits.

The PDF for $\nu_{\rm s}\sin i (R/10\,{\rm km})$ is asymmetric toward
zero and hence prohibits setting a meaningful lower limit due to the
instrinsic width of the line set by its fine structure. Had we ignored
the instrinic shape of the line and had adopted a black line profile,
which is a narrow line with zero flux about line center and fixed EW =
0.1 \AA\ (dashed line), we would set misleading constraints on the
neutron star parameters.  Using this black line profile, we can
constrain $\nu\sin i (R/10\,{\rm km}) = 47-102$ Hz which for a 44.7 Hz
rotator gives $R\sin i = 10.4 - 23$ km.  These misleading constraints
highlight the importance of using the intrinsic line profile to fit
for NS parameters.  The limitations of simple line profiles have also
been recognized previously by Villarreal \& Strohmayer (2004).

The 68\%, 95\% and 99\% central confidence intervals of ${\rm log}_{10}
(N_{\rm Fe, n=2}/{\rm cm^{-2}})$ are $17.69-18.05$, $17.48-18.17$, and
$17.34-18.26$ respectively.  The corresponding $N_{\rm Fe}$'s are
$1.4-3 \times 10^{20}\,{\rm cm}^{-2}$, $0.9-4.0 \times 10^{20}\,{\rm
cm}^{-2}$, and $0.8-4.8\times 10^{20}\,{\rm cm}^{-2}$.  At 68\%
confidence, these values are within a factor of 3-6 times the solar
metallicity photospheric value, $N_{\rm Fe} \approx 5\times
10^{19}\,{\rm cm}^{-2}$ (CBW) and within a factor of 5-10 times the
accretion-spallation value of $N_{\rm Fe} \approx 3.4\times
10^{19}\,{\rm cm}^{-2}$ (BCP).  At 99\% confidence, we may require a
column a factor of up to 10 times larger.  Our estimate from this
likelihood analysis agrees with CBW's simple EW estimate.  In addition
the NLTE value of $N_{\rm Fe}$ is likely larger (London, Taam \&
Howard 1986) and may demand either supersolar metallicity or radiative
levitation (CBW).

The 68\%, 95\% and 99\% central confidence intervals of $1+z$ are
$1.342-1.348$, $1.337-1.350$, and $1.334-1.353$.  We have also
calculated the corresponding confidence intervals for $R/M$ which
gives $R = 6.592-6.670\,(M/M_{\odot})\,{\rm km}$,
$6.565-6.722\,(M/M_{\odot})\,{\rm km}$, and
$6.542-6.766\,(M/M_{\odot})\,{\rm km}$. Since we fit a rotationally
broadened profile to the data, our measurement of $1+z$ is immune to
beaming effects which Ozel \& Psaltis (2003) (also see Bhattacharyya,
Miller \& Lamb 2004) point out can skew the measurement of the
redshift if the line energy is taken from flux minimum (or
combinations of minima).

\section{Summary and Conclusions}

Using the intrinsic Fe H$\alpha$ line profiles calculated by CBW, we
calculate rotational broadened line profiles to compare to the
observed lines in \EXO\ (CPM). We show that the line profiles
generated by the simple S+D approximation accurately reproduces the
ones from a fully relativistic ray tracing method described briefly in
\S~2.
 
Assuming that the burst emission comes from the entire surface of the
NS, we constrain the Fe column in the line forming region (${\rm
log}_{10} N_{\rm Fe, n=2}/{\rm cm^{-2}} = 17.9_{-0.42}^{+0.27}$).  We
also find that the redshift is $1 + z = 1.345_{-0.008}^{+0.005}$.  The
NS spin is $\nu_{\rm s}\sin i (R/10\,{\rm km}) = 32_{-28}^{+52}\,{\rm
Hz}$ for a fiducial NS radius of $10$ km.  These errors denote 95\%
confidence intervals. Fixing $\nu_{\rm s} = 44.7$ Hz (Villarreal \&
Strohmayer 2004), we find that the radius is effectively unconstrained
by the observed line due to the intrinsic width set by fine structure
splitting, an effect missed by simple line profiles. 

Other transitions would simultaneously confirm the redshift
measurement and may provide meaningful constraints on the radius.  CBW
calculated the line profiles for the associate Ly$\alpha$ transition
and P$\alpha$ transition.  The EWs of the Fe Ly$\alpha$ and P$\alpha$
lines were $\approx 15-25$ eV and $\approx 7-10$ eV, respectively.
The P$\alpha$ transition is obscured by interstellar absorption. CBW
points out that the Ly$\alpha$ line profile is dominated by rotational
broadening.  Hence a sufficiently deep observation of the Ly$\alpha$
transition would provide a good constraint on the radius of EXO
0748-676.  The helium-like Fe $n=2\rightarrow 3$ may provide some
additional constraints on EXO 0748-676, but the intrinsic line profile
is not known.  This is critical in modeling this feature, but is
beyond the scope of this work.  Spectral modeling of the continuum
emission during Type I x-ray burst has also been utilized to determine
NS parameters (Shaposhnikov \& Titarchuk 2004; Shaposhnikov,
Titarchuk, \& Haberl 2003 and references therein). Redshift
measurements and spectral modeling utilized in combination would set
very tight constraints on NS parameters.

\begin{figure}
\epsscale{1.0} \plotone{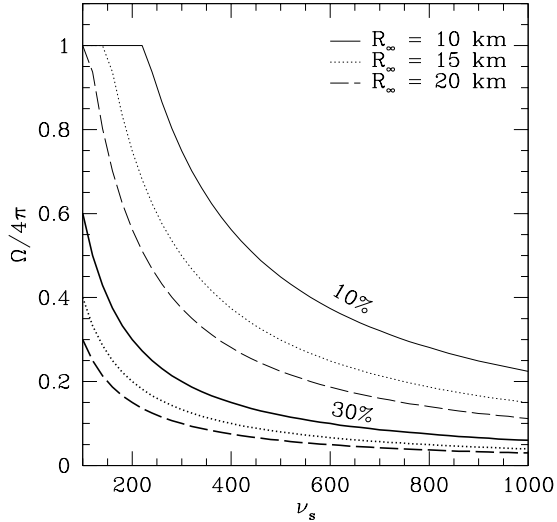}
  \caption{Fraction of Sky for which the line minimum is below 10\%
  (thin lines) and 30\% (thick lines) below the continuum as a
  function of spin. We fix the Fe H$\alpha$ line to the
  observed EW from EXO 0748-676 and plot the line for NS $R_{\infty}
  =$ 10 (solid lines), 15 (dotted lines) and 20 km (dashed lines).}
  \label{fig:detection}
\end{figure}

Though EXO 0748-676 is rotating slowly among LMXBs, more rapid
rotation does not necessarily make spectral lines undetectable. If a
rapidly rotating NS is viewed more face on, the spectral lines will be
narrower.  Hence, more rapid rotators may just constitute a smaller
population of NSs with detectable features.  We illustrate this in
Figure~\ref{fig:detection} where we plot the percentage of sky denoted
by $\Omega/4\pi$ over which the line minimum exceeds 10\% and 30\%
below the continuum. We assume the same intrinsic line profile as the
observed H$\alpha$ line on EXO 0748-676 for a NS $R_{\infty}$ of 10,
15 and 20 km. We use $R_{\infty}$ instead of $R$ because $R_{\infty}$
is a convenient way of absorbing the factor of $1 + z$ in the scale of
broadening, $\beta$.  These lines follow a very simple trend, which we
now derive.  Up to a certain rotation rate, these lines are always
detectable.  For sufficiently large rotation rates, the FWHM of the
line is $\propto \nu_{\rm s}\sin i$.  Since the EW is conserved, the
line minimum is $\propto (\nu_{\rm s} \sin i)^{-1}$.  Therefore,
maximum inclination for line minima greater than a certain strength is
$\sin i_{\rm max} \propto \nu_{\rm s}^{-1}$, which limits the fraction
of sky over which this line would be detectable.  Hence, the larger
effect of rotational broadening with higher inclination broadens the
line and weakens line minima unless emission is restricted to the
rotational pole (Ozel \& Psaltis 2003; Bhattacharyya, Miller, \& Lamb
2004).  \

For a fiducial value of $R_{\infty} = 10-15$ km, 10-20\% of NS with
spins of 300 Hz would have spectral features similar to EXO 0748-676
where the line minimum is about 30\% below the continuum.  At 600 Hz,
the percentage is reduced by a factor of two.  For these favorably
spinning and oriented NSs (line minimum $\approx 30$\%), the line
profile is set by both rotational broadening and fine structure
splitting.  Therefore, burst studies of NSs with spins near 300 Hz may
yield at least another example of atomic spectral lines from NSs.  For
these most favorable systems, the intrinsic line profile is absolutely
necessary in order to model these lines and perform meaningful
measurements from the observations.  Of the presently known LMXBs with
X-ray bursts, a very promising target is 4U 1728-34 which spins at 363
Hz (Galloway et al 2003) and exhibits bursts at a fairly regular rate.
A recent model of the NS-accretion disk geometry (Shaposhnikov et
al. 2003) estimates $i$ to be around 50$^{\circ}$. At this
inclination, the line minimum is 7\% below the continuum.  For smaller
inclinations of $i = 30^{\circ}$ and $10^{\circ}$, the line minimum
would be 10\% and 26\% respectively.

\acknowledgements

We thank Jean Cottam for providing the observational data and for
useful discussions.  We also thank the anonymous referee who comments
greatly improved this work. This work was supported by the National
Science Foundation (NSF) under PHY 99-07949, and by the Joint
Institute for Nuclear Astrophysics through NSF grant PHY 02-16783.
Support for this work was also provided by NASA through Chandra Award
Number GO4-5045C issued by the Chandra X-ray Observatory Center, which
is operated by the SAO for and on behalf of NASA under contract
NAS8-03060. S.M.M is supported by a grant from NSERC. I.W. receives
partial support from NSF grant AST-0307273. P.C. acknowledges support
from the Miller Institute for Basic Research in Science, Univsersity
of California Berkeley.

\end{document}